\documentclass[aps,prd,10pt,superscriptaddress,nofootinbib,showkeys,twocolumn]{revtex4-2}

\usepackage[T1]{fontenc}
\usepackage{lmodern}
\usepackage[scaled=1]{XCharter}
\usepackage[scaled=.95]{beramono}
\usepackage[charter,ncf,scaled=1.05]{newtxmath}
\usepackage[bb=dsserif,bbscaled=1.15]{mathalpha}

\usepackage{mathtools,siunitx,physics}
\usepackage{graphicx}
\usepackage[normalem]{ulem}
\usepackage{enumerate}
\usepackage{silence}
\ExplSyntaxOn
\msg_redirect_name:nnn{siunitx}{physics-pkg}{none}
\ExplSyntaxOff
\usepackage[dvipsnames]{xcolor}
\definecolor{theme}{rgb}{0.7, 0, 0.35}
\usepackage[colorlinks, hypertexnames=false, allcolors=theme, bookmarksnumbered=true]{hyperref}
\usepackage[capitalise]{cleveref} 
\usepackage{orcidlink}
\usepackage{xspace}

\newcommand{\ie}{i.e.,\@\xspace}

\crefname{section}{Sec.}{Secs.}
\Crefname{section}{Section}{Sections}
\crefname{appendix}{App.}{Apps.}
\Crefname{appendix}{Appendix}{Appendices}
\newcommand{\Cite}[1]{Ref.~\cite{#1}}
\newcommand{\Cites}[1]{Refs.~\cite{#1}}

\DeclareMathAlphabet{\mathpzc}{OT1}{pzc}{m}{it}

\newcommand{\msf}[1]{\mathsf{#1}}

\newcommand{\mk}{\mspace{1mu}}
\renewcommand{\vec}[1]{\boldsymbol{#1}}

\newcommand{\pd}{\partial}

\newcommand{\del}{\vec{\nabla}}

\newcommand{\sub}[1]{_\mathrm{#1}}

\providecommand{\vv}{}
\renewcommand{\vv}{\vec{v}}
\newcommand{\ee}{\vec{E}}
\newcommand{\bb}{\vec{B}}
\newcommand{\jj}{\vec{J}}

\graphicspath{{./}{figs/}{figures/}}

\newcommand{\cm}{C_5}
\newcommand{\cw}{C_\omega}
\newcommand{\kk}{k_\mu}
\newcommand{\ttt}{\tau}
\newcommand{\gf}{\Gamma_{\rm f}}
\newcommand{\highlight}[1]{{\color[RGB]{205,66,62} #1}}

\newenvironment{notation}[1][4em]{%
  \list{}{%
    \setlength\labelwidth{#1}%
    \setlength\labelsep{0.5em}%
    \setlength\leftmargin{\labelwidth}\addtolength\leftmargin{\labelsep}%
    \setlength\itemsep{2pt}\setlength\parsep{0pt}\setlength\topsep{4pt}%
  }%
}{\endlist}

\begin{document}

\title{Relativistic chiral MHD with application to the early Universe}

\author{Deepen~Lark\,\orcidlink{0000-0001-5226-1913}}
\email{deepenlark@uni-bonn.de}
\affiliation{Argelander-Institut für Astronomie, Universität Bonn, Bonn 53121, Germany}

\author{Jennifer~Schober\,\orcidlink{0000-0001-7888-6671}}
\email{schober@uni-bonn.de}
\affiliation{Argelander-Institut für Astronomie, Universität Bonn, Bonn 53121, Germany}

\date{\today}

\keywords{chiral anomaly, chiral magnetic effect, cosmological magnetic fields, relativistic plasmas}

\begin{abstract}
We present a systematic derivation of the equations of relativistic chiral magnetohydrodynamics (MHD) for a plasma of charged fermions in an expanding universe.
Through a combination of a coordinate transformation and a rescaling of the dynamical variables, we bring the full system to the same form as in the Minkowski metric, with the Hubble expansion surviving only in the chirality-flipping rate and the kinematic viscosity.
Retaining all four contributions to both the electric and the axial current yields terms absent from standard chiral MHD: charge-density corrections to the evolution equations for the chemical potentials, and an electric current proportional to the charge chemical potential $\mu$ and the bulk velocity.
The latter is mandated by current conservation, requires no chirality imbalance, and drives the charge-flow instability studied in a follow-up paper. 
For the radiation-dominated era, we evaluate all the coefficients in physical units and use them to estimate the magnetic Reynolds number, the attainable magnetic field strength, and the minimum temperature at which the chiral dynamo can operate.
The resulting equations are cast in a form ready for direct numerical implementation.
\end{abstract}

\maketitle




\section{Introduction}
\label{sec:intro}

Magnetic fields pervade the Universe, having been observed in planets \cite{ref:schubert11}, stars \cite{ref:donati09,ref:solanki06}, galaxies \cite{ref:beck15}, galaxy clusters \cite{ref:carilli02}, and perhaps even in the voids of the large-scale structure \cite{ref:neronov10b}, yet their origin in many of these systems remains an open problem \cite{ref:roberts13,ref:charbonneau14,ref:charbonneau20,ref:brandenburg05,ref:durrer13}.
On cosmological scales, magnetic fields have been invoked to explain the spectra of TeV blazars \cite{ref:neronov09,ref:neronov10a,ref:neronov10b}, the non-observation of secondary GeV cascade emission implying a lower bound on the field strength in the intergalactic voids.
Such an explanation requires a high volume-filling factor \cite{ref:dolag11,ref:tjemsland24}, suggesting that a primordial origin is more likely than an astrophysical one \cite{ref:beckam13,my:dipole,ref:ghosh26,ref:seller26,ref:hosking23}.
Most primordial magnetogenesis scenarios invoke physics beyond the Standard Model, operating either during inflation \cite{ref:turner88,ref:ratra92,ref:kunze10,ref:anber10,ref:barnaby11,my:inf_noback,tex:gerlach26} or in cosmological first-order phase transitions  \cite{ref:hogan83,ref:quashnock89,ref:vachaspati91,ref:baym96,ref:sigl97,ref:ellis19}.
Despite decades of effort \cite{ref:durrer13,ref:batista21,ref:grasso01,ref:kandus11}, however, a viable mechanism free of significant
fine-tuning is yet to be established.

These primordial fields, if present, would be a rare cosmological
messenger. As surviving relics, they would offer a direct probe of the optically thick epochs before recombination, which are inaccessible to electromagnetic observations.
Their presence has been invoked, for example, to relax the Hubble tension \cite{ref:jedamzik26}, to help alleviate the lithium abundance problem \cite{ref:yamazaki14}, and to modify and source primordial gravitational waves \cite{ref:roperpol22,ref:caprini09} (see \Cites{ref:durrer13,ref:subramanian16,ref:vachaspati21} for detailed reviews).
They could also provide the seed fields for galactic dynamos, an
option that appears increasingly necessary as coherent magnetic fields are observed in galaxies at ever higher redshifts \cite{ref:mao17,ref:geach23,ref:deroo25,ref:chen24},
leaving progressively less time for a dynamo to amplify weak astrophysical seeds.

Magnetogenesis during first-order phase transitions at the
electroweak (EW) and QCD scales has also been linked to baryogenesis,
since the magnetic helicity constitutes the electromagnetic (EM) part
of the Chern--Simons number \cite{ref:fujita16, ref:vachaspati01, tex:boyer25}.
This would couple the generation of helical fields to baryogenesis, with the latter providing the fermions with net nonzero chemical potential $\mu$.
If the process responsible for a nonzero $\mu$ also produces a nonzero chiral chemical potential $\mu_5$ \cite{ref:campbell92,ref:kharzeev11,ref:burnier11,ref:domcke23, ref:gurgenidze26}, both being specific linear combinations of the chemical potentials of the individual fermion species of given charge and chirality, then chiral effects would actively shape the subsequent evolution of cosmological magnetic fields \cite{ref:joyce97,ref:boyarsky12,ref:rogachevskii17,ref:schober24prl,ref:schober24prd}.

A nonzero $\mu_5$ can also arise in other settings: chiral effects have been studied in early-Universe processes such as the generation of chiral gravitational waves \cite{ref:okano21,ref:brandenburg24,ref:kahniashvili21,tex:gerlach26}, as well as in heavy-ion collisions \cite{ref:kharzeev24,ref:kharzeev16}, condensed-matter systems \cite{ref:son13,ref:huang15,ref:xiong15,ref:li16}, magnetars \cite{tex:ohnishi14, ref:sigl16, tex:dehman24, ref:dehman25, ref:dehman26}, core-collapse supernovae \cite{ref:yamamoto16, ref:grabowska15, ref:masada18, ref:matsumoto22}, and pulsar kicks \cite{ref:kaminski16,ref:charbonneau10}.
In relativistic plasmas, the presence of a chirality imbalance can induce additional currents through a quantum anomaly, namely the chiral anomaly.

The chiral anomaly is the nonconservation of the axial current of massless Dirac fermions in the presence of gauge fields \cite{ref:adler69,ref:bell69}. Owing to its topological nature, the anomaly coefficient is exact and survives at macroscopic scales.
Resultantly, the anomaly endows relativistic plasmas with transport phenomena that have no counterpart in classical magnetohydrodynamics (MHD) and that manifest as additional electric currents (for charged fermions) and axial currents.
These additional currents alter the evolution of magnetic fields.
For example, the chiral magnetic effect (CME) \cite{ref:vilenkin80,ref:fukushima08,ref:rybalka19} can amplify existing magnetic fields, while the chiral vortical effect (CVE) \cite{ref:vilenkin79,ref:son09,ref:huang19} can even generate them from vorticity and amplify them further, either directly, in the case of charged fermions \cite{ref:wang24}, or  indirectly, through hydrodynamic interactions, in the case of neutral fermions such as neutrinos \cite{ref:yamamoto16}.
Understanding chiral MHD is thus necessary to pinpoint the evolution of magnetic fields in these systems.

While chiral MHD has been studied extensively over the past few decades, most of this work has focused on the effects of individual anomalous terms, manifesting, for example, as the chiral plasma instability \cite{ref:vilenkin80,ref:joyce97,ref:boyarsky12,ref:rogachevskii17}, the chiral magnetovortical instability \cite{ref:wang24,ref:das25,tex:wang25,ref:wang26}, or chiral magnetic waves \cite{ref:kharzeev11,ref:ikeda23}.
A more rigorous formulation of chiral hydrodynamics, derived from chiral kinetic theory with dissipative effects included and Lorentz covariance retained, was presented in \Cite{ref:rybalka19}.
In dynamo studies, the anomalous currents are typically appended to the familiar nonrelativistic MHD equations, in order to focus on uncovering the rich phenomenology of chiral dynamos \cite{ref:schober24prl,ref:schober24prd,ref:schober22,ref:schober20,ref:schober18,ref:rogachevskii17,ref:brandenburg17,ref:masada18,ref:brandenburg21b}.
It is therefore useful to obtain the dynamo-ready equations of chiral MHD directly from a covariant formulation, so that the set of
retained terms and the relations among their coefficients follow from the derivation itself rather than being supplied term by term.

Here, we present a systematic derivation of the equations of chiral MHD for a plasma comprising relativistic charged fermions, with a view to early-Universe applications.
We first show explicitly that, as for ideal MHD \cite{ref:brandenburg96}, the equations of chiral MHD in an expanding universe can be brought to their flat-spacetime form by a suitable transformation of coordinates and rescaling of the dynamical variables.
We then derive these equations in the 1+3 formalism, starting from a covariant formulation and adopting the one-fluid approximation.
We also evaluate all the coefficients in physical units, identifying which of them can be treated as independent and which are related to one another.
This provides a common reference for comparing results across the literature \cite{ref:long16,ref:brandenburg21b,ref:brandenburg24,ref:gurgenidze26,ref:boyarsky21,ref:boyarsky21b,ref:rogachevskii17,ref:gorbar16a,ref:dehman25}, where a variety of conventions are in use, and it clarifies which coefficients may be varied independently in parameter studies.

While many of the terms arising in this derivation can be dropped on the basis of order-of-magnitude estimates, a few cannot be discarded a priori; some of these are absent from the standard chiral MHD equations \cite{ref:rogachevskii17,ref:schober24prd,ref:gurgenidze26}.
Notably, one of these terms is an electric current that requires no chirality imbalance.
It drives an additional instability, which we
call the charge-flow (C-flow) instability, and under certain conditions, it can even exceed the CME current.
We study this instability in detail in our follow-up paper \cite{my:cflow_theory}.
In line with standard one-fluid MHD, we neglect the displacement current here.
However, as we show below (\cref{eq:mainchiraleqs}) and discuss further in \Cite{my:cflow_theory}, a dynamical evolution of the electric field may ultimately be needed, which would likely require a two-fluid treatment; we defer this to future work.

In \cref{sec:prelim}, we demonstrate how the equations of chiral MHD in an expanding universe can be brought to the same form as in the Minkowski metric.
While the procedure largely parallels that for ideal MHD, our presentation aims to clarify the distinction among the various coordinate transformations and rescalings involved.
If the chiral dynamics are instead studied in Minkowski space, \cref{sec:prelim} can be skipped altogether and the scale factor simply set to unity throughout.
In \cref{sec:chiral}, we derive the equations of chiral MHD in the 1+3 formalism. We then discuss the limits that are motivated by the conditions in the early Universe \cite{ref:domcke23}, although the derivation itself remains general and does not presuppose specific parameter values.
In \cref{sec:early}, we turn to the application to the early Universe and obtain a set of equations that can be used directly for analytical and numerical MHD treatments.
In \cref{sec:concl}, we summarize our results and present an outlook.
Finally, \cref{sec:abbr} lists the abbreviations and symbols used in this work.

\section{Preliminaries}
\label{sec:prelim}

\subsection{Relativistic ideal MHD in an expanding universe}

We use the spatially flat Friedmann--Lemaître--Robertson--Walker (FLRW) metric given by
\begin{gather}
    \dd s^2 = a^2 \qty(-\dd \ttt^2 + \delta_{ij} \dd x^i \dd x^j) \,,
    \label{eq:metric}
\end{gather}
where $\ttt$ is the conformal time, and $a = a(\ttt)$ is the scale factor. Herein, Greek indices are used to denote the spacetime components \([0,1,2,3] \equiv [\ttt,x,y,z] \) and Latin indices are used for the spatial components \([1,2,3] \equiv [x,y,z]\).  We use $\equiv$ to denote definitions. We assume units such that \(c = k\sub{B} = 1\) but we retain \(\hbar\) and the elementary charge \(e\) explicitly to facilitate later calculations linking the theory to numerical setups and
observations.

Denoting the EM field tensor as $F_{\alpha\beta}$, Maxwell's equations read
\begin{gather}
    \nabla_\beta F^{\alpha \beta} = \mu_0 J^\alpha \,,
    \qquad \frac{1}{2}
    \mk \epsilon^{\alpha\beta\gamma\delta} \mk\mk \nabla_\beta
    F_{\gamma\delta}
    = 0 \,,
    \label{eq:max}
\end{gather}
where \(\mu_0\) is the vacuum permeability, 
\(J^\alpha\) is the four-current defined to be a proper tensor, \(\epsilon_{\alpha\beta\gamma\delta}\) is the totally antisymmetric Levi-Civita tensor with \(\epsilon_{0123} = \sqrt{-g}\), $g$ is the determinant of the FLRW metric $g_{\alpha\beta}$, and we use \(\nabla_\alpha\) to denote covariant derivative along the tangent vector \(\pd_\alpha\) with Christoffel symbols used as the affine connection.

The electric and magnetic fields are defined with respect to a comoving observer with velocity 
\(n^\alpha\), which is equal to \(\qty(1/a,0,0,0)\) in the FLRW coordinates \eqref{eq:metric},
\begin{gather}
    E^\alpha \equiv F^{\alpha\beta} n_\beta \,,
    \qquad
    B^\alpha \equiv - \frac{1}{2}
    \mk \epsilon^{\alpha\beta\gamma\delta} \mk 
    n_\beta \mk F_{\gamma\delta} \,.
\end{gather}
If we perform a coordinate transformation, \(\dd x^\alpha \to \dd \bar{x}^\alpha = a \dd x^\alpha\), such that \(g_{\alpha\beta} \to \bar g_{\alpha\beta} = \eta_{\alpha\beta}\), where \(\eta_{\alpha\beta}\) is the Minkowski metric, any vector \(X^\alpha \to \bar X^\alpha = a X^\alpha\) and any one-form \(X_\alpha \to \bar X_\alpha = X_\alpha/a\). These transformed quantities, denoted by an overbar, can be understood as the physical measurements made by a local observer. Thus, we can write the comoving electric and magnetic fields in the FLRW coordinates \eqref{eq:metric} in terms of these physical quantities as
\begin{gather}
    E_\alpha = a \bar E_\alpha = a(0, \bar{\ee}) \,,
    \qquad
    B_\alpha = a \bar B_\alpha = a(0, \bar{\bb}) \,.
\label{eq:ebbar1}
\end{gather}
Herein, we use boldface to denote three-dimensional spatial variables in the index-free form. Then, Maxwell's equations \eqref{eq:max} for the comoving fields in the FLRW coordinates \eqref{eq:metric} can be written as
\begin{equation}
\begin{gathered}
    \pd_\ttt \qty(a^2\bar{\bb}) + a^2 \del \times \bar\ee = 0 \,,
    \\
    a^2 \mk \del \times \bar\bb = \pd_\ttt \qty(a^2 \mk\bar\ee) + \mu_0 \mk a^3 \bar{\vec J} \,,
    \\
    \del \cdot \bar\bb = 0 \,,
    \qquad \del \cdot \bar\ee = a \mk \mu_0 \mk \bar\rho_{\rm el} \,,
\end{gathered}
\label{eq:max_unsc}
\end{equation}
where we introduce \(\bar\rho_{\rm el} \equiv -\bar J^\alpha \bar n_\alpha = \bar J^0\), \(\bar{\vec{J}} \equiv \bar J^i\), and \(\cdot\) and \(\times\) denote the dot product and the cross product of the respective three-vectors using Euclidean metric. Here, \(\del \equiv \pd_{\vec x}\) is the usual three-dimensional del operator, and the spatial and time derivatives are with respect to the coordinates introduced in \cref{eq:metric}, even though the spatial vector fields from the Minkowski space are used to write the original four-vector fields.

Equations~\eqref{eq:max_unsc} can be brought to the same form as in the Minkowski space
\begin{subequations}
\begin{gather}
    \pd_\ttt \widetilde{\bb} + \del \times \widetilde\ee = 0 \,,
    \label{eq:farday_sc}
    \\
    \del \times \widetilde\bb = \pd_\ttt \widetilde\ee + \mu_0 \mk \widetilde{\vec J} \,,
    \label{eq:ampere_sc}
    \\
    \del \cdot \widetilde\bb = 0 \,,
    \qquad \del \cdot \widetilde\ee = \mu_0 \mk\mk \widetilde\rho_{\rm el} \,,
    \label{eq:gauss_sc}
\end{gather}
\label{eq:max_sc}
\end{subequations}
by rescaling the different quantities,
\begin{equation}
\begin{gathered}
    \widetilde \bb = a^2 \mk \bar\bb \,,
    \quad \widetilde \ee = a^2 \mk \bar\ee \,,
    \quad \widetilde \jj = a^3 \mk \bar\jj \,,
    \\
    \quad \widetilde \rho_{\rm el} = a^3 \mk \bar \rho_{\rm el} \,,
    \quad \widetilde n^\alpha = \bar n^\alpha \,,
\end{gathered}
\label{eq:scaling1}
\end{equation}
where we use tilde to denote the rescaled quantity. Note that while the three-vectors are scaled as in the equation above, the corresponding four-vectors have one additional factor of $a$ and the corresponding one-forms have one less, which arises from the coordinate transformation from the FLRW to the Minkowski metric, for example,
\begin{equation}
\begin{gathered}
    \widetilde B^\alpha = a^3 B^\alpha \,,
    \quad \widetilde E_\alpha = a E_\alpha \,,
    \quad \widetilde J^\alpha = a^4 J^\alpha \,,
    \\
    \quad \widetilde \rho_{\rm el} = a^3 \rho_{\rm el} \,,
    \quad \widetilde n^\alpha = a n^\alpha \,,
    \quad \widetilde n_\alpha = \frac{1}{a} n_\alpha \,.
\end{gathered}
\label{eq:scaling3}
\end{equation}

It is also useful to define the electric field and the magnetic field observed by a fluid element, which are also the fields that go into the Lorentz force law and the constitutive relations for MHD,
\begin{gather}
    e^\alpha \equiv F^{\alpha\beta} u_\beta \,,
    \qquad
    b^\alpha \equiv - \frac{1}{2}
    \mk \epsilon^{\alpha\beta\gamma\delta} \mk
    u_\beta \mk F_{\gamma\delta} \,,
    \label{eq:emfluid}
\end{gather}
where \(u^\alpha = \gamma (1,\vec{v})/a\) is the bulk velocity of the fluid element in the comoving coordinates, \(\vec{v}\) is the corresponding 
three-velocity, and 
the Lorentz factor is  \(\gamma = \left(1 - v^2\right)^{- 1/2}\). The expression for the components of \(e^\alpha\) and \(b^\alpha\) in the fluid's non-expanding rest frame in terms of \(\bar{\ee}\) and \(\bar{\bb}\) can be found in, for example, Ref.~\cite[Eqs.~(11.149)]{book:jackson}. Maxwell's equations in terms of \(e^\alpha\) and \(b^\alpha\) can be found in, for example, Refs.~\cite[App.~B]{ref:durrer13} or \cite[Sec.~15.3]{book:shukurov}. However, we will continue using the rescaled Maxwell's equations~\eqref{eq:max_sc}. To that end, let us also write the rescaled \(e^\alpha\) and \(b^\alpha\) in the FLRW coordinates \eqref{eq:metric} in terms of the fields that go into Eqs.~\eqref{eq:max_sc},\footnote{Note that the fields in Eqs.~\eqref{eq:eb_small} are not simple Lorentz transformations of \(E^\alpha\) and \(B^\alpha\) to the fluid's rest frame, as can be checked directly by transforming \cref{eq:ebbar1} into a moving frame with three-velocity \(\vv\).}
\begin{equation}
\begin{aligned}
    \widetilde b^\alpha
    &= \left(\gamma \mk\mk \vec{v} \cdot \widetilde\bb,
     \; \gamma \mk \widetilde \bb - \gamma \mk\mk \vec{v} \times \widetilde \ee \right) \,,
    \\
    \widetilde e^\alpha
    &= \left(\gamma \mk\mk \vec{v} \cdot \widetilde \ee,
     \; \gamma \mk \widetilde \ee + \gamma \mk\mk \vec{v} \times \widetilde \bb\right) \,.
\end{aligned}
\label{eq:eb_small}
\end{equation}
All the explicit components of tensors in this article are written in the comoving FLRW coordinates \eqref{eq:metric}.

During the radiation era, if we rescale the total energy density
\begin{gather}
    \widetilde \rho = a^4 \bar  \rho \,,
    \label{eq:rhoscaling}
\end{gather}
it can be shown \cite{ref:brandenburg96} that, in addition to Maxwell's equations, the hydrodynamic equations also take the same form as their Minkowski version if the bulk velocity \(u^\alpha\) scales like \(n^\alpha\),
\begin{gather}
    \widetilde u^\alpha = a \mk u^\alpha
    \quad \to \quad \widetilde \vv = \vv \,.
\end{gather}
Moreover, the rescaled vorticity observed by the fluid element can be written in the FLRW coordinates as
\begin{gather}
    \widetilde \omega^\alpha \equiv
    -a^2 \epsilon^{\alpha\beta\gamma\delta} \mk u_\beta \pd_\gamma u_\delta
    = \gamma^2 \left(\vec{\omega} \cdot \vec{v},
     \vec{\omega} + \vec{v} \times  \pd_\ttt \vec{v} \right) \,,
\end{gather}
where we have defined \(\vec \omega \equiv \del \times \vv\) and we have rescaled  the vorticity tensor by \(a^2\) like the EM fields.

Using these scaled quantities and conformal time, Hubble expansion and scale parameter $a$ can thus be subsumed into the Minkowski-space ideal MHD equations, making the solution for these rescaled quantities completely independent of $a$.

\subsection{Chiral (resistive) MHD in an expanding universe}

While the previous subsection was concerned with ideal MHD, here we reinstate the ohmic current, which is parallel to \(e^\alpha\). Moreover, the chiral anomaly in the Standard Model leads to additional electric currents in the direction of the magnetic field and the vorticity observed by the fluid element. The axial charge obeys an analogous transport structure. Both the electric current and the axial current contain four contributions each, one longitudinal term related to the respective charge flow and three transverse terms that are proportional to the electric field, the magnetic field and the vorticity, respectively. The electric current is
\begin{gather}
    J^\alpha = J^\alpha_{\rm flow} + J^\alpha_{\rm ohm} + J^\alpha_{\rm CME} + J^\alpha_{\rm CVE},
    \label{eq:j}
\end{gather}
and the individual currents are given by \cite{ref:rybalka19, ref:isachenkov11}
\begin{subequations}
\begin{align}
    J^\alpha_{\rm flow} &= \rho_{\rm el} u^\alpha \,,
    &J^\alpha_{\rm ohm} &= \sigma_E e^\alpha \,,
    \label{eq:jterm1}
    \\
    J^\alpha_{\rm CME} &= \frac{e^2}{2 \pi^2 \hbar^2} \mu_5 T b^\alpha \,, \!\!
    &J^\alpha_{\rm CVE} &= \frac{e}{\pi^2 \hbar^2} \mu \mu_5 T^2\omega^\alpha \,,
    \label{eq:jterm2}
\end{align}
\end{subequations}
where $\sigma_E$ is the ohmic conductivity of the plasma, \(T\) is the temperature of the plasma, and \(\mu\) and \(\mu_5\) are the dimensionless plasma chemical potential and the chiral chemical potential, respectively, both normalized by \(T\),
\begin{gather}
    \mu = \frac{\mu^{\mathrm{phys}}}{T} \,,
    \qquad \mu_5 = \frac{\mu_5^{\mathrm{phys}}}{T} \,,
    \label{eq:mu_norm}
\end{gather}
where \(\mu^{\mathrm{phys}}\) and \(\mu^{\mathrm{phys}}_5\) are the corresponding physical potentials with units of energy. In \cref{eq:jterm1}, the first term is the current associated with the flow of the electric charge; the second term is the ohmic current. In \cref{eq:jterm2}, the first term is the CME current \cite{ref:vilenkin80,ref:fukushima08}; the second term is the CVE current \cite{ref:vilenkin79,ref:son09,ref:huang19,ref:wang24}.

Similarly, the axial current can be written as~\cite{ref:rybalka19}
\begin{gather}
    J^\alpha_5 = J^\alpha_{\rm 5,\,flow} + J^\alpha_{\rm CSE} + J^\alpha_{\rm CESE} + J^\alpha_{\rm AVE} \,,
\end{gather}
with individual terms given by
\begin{subequations}
\begin{align}
    J^\alpha_{\rm 5,\,flow} &= \rho_5 u^\alpha \,,
    \enspace
    J^\alpha_{\rm CESE} = \frac{6 \mu \mu_5}{\pi^2 + 3\mu^2 + 3 \mu_5^2} \sigma_E \mk e^\alpha \,,
    \label{eq:j5term1}
    \\
    J^\alpha_{\rm CSE} &= \frac{e^2}{2 \pi^2 \hbar^2} \mu \mk T \mk b^\alpha \,,
    \label{eq:j5term2}
    \\
    J^\alpha_{\rm AVE} &= \frac{e}{6\pi^2 \hbar^2} T^2 \left(
    \pi^2 + 3\mu^2 + 3\mu_5^2
    \right)
    \omega^\alpha \,,
    \label{eq:j5term3}
\end{align}
\end{subequations}
where \(\rho_5 = -J_5^\alpha n_\alpha\) is the proper density of the axial charge with the same units as \(\rho_{\rm el}\) (both equal the respective number densities multiplied by $e$). The first term in \cref{eq:j5term1} is the axial current associated with the flow of the axial charge density, and the second term is the current associated with the chiral electric separation effect (CESE) \cite{ref:huang13}. Note that the latter is much smaller than the ohmic current in the limit of \(\mu, \mu_5 \ll 1\). While \cref{eq:j5term2} denotes the axial current due to the chiral separation effect (CSE) \cite{ref:lin18}, the term in \cref{eq:j5term3} is the axial current due to the axial vortical effect (AVE) \cite{ref:deng16,ref:lin18}.

In the hydrodynamic limit, the electric and the axial charge densities can be expressed in terms of the chemical potentials (a detailed derivation can be found in \Cite{ref:rybalka19}),
\begin{equation}
\begin{aligned}
    \rho_{\rm el} &= \mu\mk e\mk T^3 \frac{\mu^2 + 3\mu_5^2 + \pi^2}{3 \pi^2 \hbar^3},
    \\
    \rho_5 &= \mu_5\mk e\mk T^3 \frac{\mu_5^2 + 3\mu^2 + \pi^2}{3 \pi^2 \hbar^3}.
\end{aligned}
\label{eq:chargemu}
\end{equation}
In order to rescale the electric current and the axial current consistently with no explicit dependence on the scale factor, all their components must be scaled uniformly, from which we obtain
\begin{equation}
\begin{gathered}
    \widetilde \mu_5 = \mu_5 \,,
    \quad \widetilde \mu = \mu \,,
    \quad \widetilde T = a T \,,
    \quad \widetilde \sigma_E = a \sigma_E \,,
    \\
    \quad \widetilde \rho_5 = a^3 \rho_5 \,,
    \quad \widetilde J^\alpha_5 = a^4 J^\alpha_5 \,.
\end{gathered}
\label{eq:scaling2}
\end{equation}
This rescaling allows the MHD equations for the magnetic field, velocity and energy density in terms of the rescaled quantities to be independent of the scale factor. In order for the whole system to be independent of the scale factor, we require the corresponding equations for \(\mu\) and \(\mu_5\) to be independent of the scale factor as well.

The equations for \(\mu\) and \(\mu_5\) can be obtained from the conservation of \(J^\alpha\),
\begin{gather}
    0 = \nabla_\mu J^\mu
    = \frac{1}{a^4} \pdv{x^\mu} \qty(a^4 J^\mu)
    \quad \to \quad
    \pd_\mu \widetilde J^\mu = 0\,,
    \label{eq:mu0}
\end{gather}
and the anomalous nonconservation of \(J^\alpha_5\) with the decay term due to chirality flipping \cite{ref:skoutnev26,ref:campbell92,ref:bodeker19,ref:boyarsky21,ref:boyarsky21b} and no extra source term,
\begin{align}
    \nabla_\mu J_5^\mu
    &= -\gf \rho_5
    + \frac{e^3}{2 \pi^2 \hbar^2} E_\mu B^\mu
    \nonumber \\
\quad \to \quad
    \pd_\mu \widetilde J_5^\mu
    &= - \widetilde \Gamma_{\rm f} \mk \widetilde \rho_5
    + \frac{e^3}{2 \pi^2 \hbar^2} \widetilde E_\mu \widetilde B^\mu\,
    \label{eq:mu50},
\end{align}
where the rescaled flipping rate is calculated from the physical value of the flipping rate as
\begin{align}
    \widetilde \Gamma_{\rm f} &\equiv a \gf^{\rm phys} \,.
    \label{eq:gammaf}
\end{align}
If \(\widetilde \Gamma_{\rm f}\), $\widetilde T$ and $\widetilde \sigma_E$ are independent of $a$, then the equations for both \(\mu\) and \(\mu_5\) are independent of $a$, in which case the full set of resistive chiral MHD equations can also be written in terms of the rescaled quantities and conformal time in the same form as their Minkowski counterparts with the effect of the Hubble expansion subsumed. However, this fails to be true for some epochs in the early Universe. For example, the rescaled flipping rate \(\widetilde \Gamma_{\rm f}\) is not independent of $a$ for temperatures below the electroweak (EW) scale \cite{ref:campbell92,ref:bodeker19,ref:boyarsky21,ref:boyarsky21b}. 
We still scale the physical \(\gf^{\rm phys}\) according to \cref{eq:gammaf} and note that the dependence of \(\widetilde \Gamma_{\rm f}\) on $a$ will make the chiral dynamics dependent on the temperature scale, which is ultimately how chiral effects vanish at lower temperatures in this description. Also note that we have reinstated only resistive and chiral effects in the ideal MHD setup. However, adding viscosity to the momentum equation can also potentially break this scaling, on which we will comment when we explicitly write the momentum equation \eqref{eq:mom}.

Hereafter, we will drop the tilde from all the quantities for brevity and assume implicitly that all quantities are rescaled as per \cref{eq:scaling1,eq:scaling2,eq:gammaf}. The one exception is the temperature: to avoid confusion, we denote the rescaled temperature by $T_0  \equiv \widetilde T = aT$ and reserve $T$ for the physical temperature. For the early-Universe application, $T_0$ is the present-day temperature of the CMB.

\section{Equations governing chiral dynamics}
\label{sec:chiral}

\subsection{Subrelativistic bulk velocity}

With this setup, we can write the equations governing the plasma dynamics in the presence of the chiral anomaly. 
We start with the spatial components of \cref{eq:j},
\begin{multline}
    \vec{J}
    =  \frac{e\mk T_0^3}{3 \hbar^3} \mu \left(1+ \frac{\mu^2 + 3\mu_5^2}{\pi^2} \right) \gamma \mk \vec{v}
    + \frac{1}{\mu_0 \eta} \gamma \left(\vec{E} + \vec{v} \times \vec{B}\right)
    \\
    + \frac{e^2\mk T_0}{2 \pi^2 \hbar^2} \mu_5 \mk
     \gamma \left(\vec{B} -\vec{v} \times \vec{E} \right)
    \\
    + \frac{e\mk T_0^2}{\pi^2\hbar^2} \mu\mk \mu_5\mk 
    \gamma^2 \left( \vec{\omega} + \vec{v} \times  \pd_\ttt \vec{v} \right) \,,
    \label{eq:jj}
\end{multline}
where \(\eta \equiv 1/(\mu_0\sigma_E)\) is the magnetic diffusivity. Let us assume subrelativistic bulk velocity such that
\begin{gather}
    \gamma \approx 1 \;\Leftrightarrow\; v^2 \ll 1 \,,
    \qquad
    \norm{\vec{v} \times \vec{E}} \ll \norm{\vec{B}} \,,
    \label{eq:subrel}
\end{gather}
where $\norm{\ldots}$ denotes the norm. Substituting all this into \cref{eq:ampere_sc} while neglecting the displacement current yields the expression for the electric field
\begin{multline}
    \vec{E} = - \vec{v} \times \vec{B}
    + \eta
    \left[ \del \times \vec{B}
    - \frac{4 \alpha \pi T_0^3}{3\mk e\mk\hbar^2} \mu \left(1+ \frac{\mu^2 + 3\mu_5^2}{\pi^2}\right) \vec{v}
    \right.
    \\ \left.
    - \frac{2 \alpha T_0}{\pi \hbar} \mu_5 \bb
    - \frac{4 \alpha T_0^2}{e \pi \hbar} \mu \mu_5 
     \left( \vec{\omega} + \vec{v} \times  \pd_\ttt \vec{v} \right) \right] \,,
    \label{eq:elecfield}
\end{multline}
where we denote the fine-structure constant by \(\alpha = e^2 \mu_0 /(4 \pi \hbar)\), and we neglect terms that are \(\order{v^2}\). If the displacement current is included, this equation instead provides the dynamical evolution of the electric field.
The main difference between \cref{eq:elecfield} and similar expressions reported in the literature \cite{ref:rogachevskii17,ref:schober24prd,ref:gurgenidze26} is the second term in the square brackets, proportional to \(\mu \vec{v}\).
This term can dominate previously studied effects such as the CVE (the last term in the second line); in environments 
where \(\mu \approx \mu_5\) and the bulk velocity is non-vanishing, 
it can even exceed the CME term (the first term in the second line). Notably, unlike the chiral effects, this term requires no chirality imbalance (since only a nonzero $\mu$ and a bulk flow suffice) and is therefore present even in plasmas where $\mu_5$ has fully decayed. We study its consequences in detail in the follow-up paper \cite{my:cflow_theory}, where we show that this term renders the system linearly unstable and study the resulting charge-flow (C-flow) instability.

Going further, \cref{eq:elecfield} can be substituted into \cref{eq:farday_sc} to obtain our first equation for the plasma dynamics, the induction equation,
\begin{multline}
    \frac{\pd \vec{B}}{\pd \ttt} = \del \times \left\{\vec{v} \times \vec{B}
    - \eta
    \left[\del \times \vec{B}
    - \kk\mk \mu_5\mk \bb
    \frac{}{}\right. \right.
    \\
    - \frac{2 \pi^2 T_0^2 }{3\mk e\mk \hbar} \kk\mk \mu \left(1+ \frac{\mu^2 + 3\mu_5^2}{\pi^2} \right) \vec{v}
    \\
    \left. \left.
    - \frac{2 T_0}{e} \kk\mk \mu\mk \mu_5 
    \left( \vec{\omega} + \vec{v} \times  \pd_\ttt \vec{v} \right) \right] \right\} \,,
    \label{eq:ind1}
\end{multline}
where we introduced
\begin{gather}
    \kk \equiv \frac{2 \alpha T_0}{\pi \hbar} \,,
    \label{eq:k5}
\end{gather}
which can be understood as the characteristic wavenumber of the chiral dynamics since the chiral plasma instability grows maximally at \(k_5 = \kk \mu_5/2\) in the linear stage \cite{ref:rogachevskii17,ref:schober18}.\footnote{This notation is different from some of the existing literature \cite{ref:rogachevskii17,ref:schober24prd,ref:gurgenidze26}, where the product \(\mu_5 \kk\) is itself denoted by \(\mu_5\). To recover that notation, one can replace \(\mu_5\) and \(\mu\) here with $\mu_5/\kk$ and $\mu/\kk$, respectively.} Note that the length scale corresponding to $\kk$ is roughly two orders of magnitude larger than the thermal de Broglie wavelength of photons at temperature $T_0$.

This is supplemented by the equations for \(\vec{v}\) and \(\rho\), which are unchanged by 
chiral effects \cite{ref:rogachevskii17},
\begin{gather}
    {\frac{D \vec{v}}{D \ttt}}
    = \frac{(\vec{\del} \times \vec{B}) \times \vec{B}}{\mu_0 \rho}
    - \frac{\vec{\del} p}{\rho} + \frac{\vec{\del} \cdot (2 \nu \rho \vec{\msf{S}})}{\rho} \,,
    \label{eq:mom}
    \\
    \frac{D\rho}{D \ttt} = - \rho \mk \vec{\del}  \cdot \vec{v} \,,
    \label{eq:mass}
\end{gather}
where \(D/D\ttt \equiv \pd_\ttt + \vv \cdot \del\) is the advective time derivative, \(\rho\) is the energy density of the plasma, $p$ is the pressure, \(\vec S\) is the rate-of-strain tensor, and \(\nu\) is the kinematic viscosity. From here, we see that in order for the momentum equation to be independent of the scale factor, the rescaled kinematic viscosity (which is rescaled like the magnetic diffusivity, \(\nu \equiv \widetilde \nu = \nu^{\rm phys}/a\)) has to be independent of the scale factor. This usually fails to be true for the early Universe, which introduces another point where the scale factor or the temperature scale enters the dynamics.

These equations need to be complemented by the equations for \(\mu\) and \(\mu_5\), which can be obtained by substituting \cref{eq:chargemu} into \cref{eq:mu0,eq:mu50},
\begin{widetext}
\begin{multline}
    \pd_\ttt \left[
    \mu \left(1 + \frac{\mu^2 + 3 \mu_5^2}{\pi^2}\right)
    + \frac{3e\hbar}{2 \pi^2 T_0^2\kk} \vv \cdot \qty(\del \times \bb)
    \right]
    \\
    + \del \cdot \left[
    \mu \left(1 + \frac{\mu^2 + 3 \mu_5^2}{\pi^2}\right) \vec{v}
    + \frac{3e\hbar}{2 \pi^2 T_0^2\eta \kk} \left(\vec{E} +  \vec{v} \times \vec{B} \right)
    + \frac{3e\hbar}{2 \pi^2 T_0^2} \mu_5 \bb
    + \frac{3\hbar}{\pi^2 T_0} \mu \mu_5 \left(\vec{\omega} + \vec{v} \times  \pd_\ttt \vec{v}\right)
    \right] = 0 \,,
    \label{eq:mu1}
\end{multline}
\begin{multline}
    \pd_\ttt \left[
    \mu_5\left(1 + \frac{\mu_5^2 + 3 \mu^2}{\pi^2}\right)
    + 9 \mu \mu_5 \frac{e \mk\hbar}{\kk\mk \eta\mk \pi^4\mk T_0^2} \vec{v} \cdot \vec{E}
    + \frac{3 \mk e\mk \hbar}{2 \pi^2 T_0^2} \mu\mk \vec{v} \cdot \vec{B}
    + \frac{\hbar}{2 T_0} \left(1+
    \frac{3\mu^2 + 3\mu_5^2}{\pi^2}\right) \vec{\omega} \cdot \vec{v}
    \right]
    \\
    + \del \cdot \left[
     \mu_5\left(1 + \frac{\mu_5^2 + 3 \mu^2}{\pi^2}\right) \vec{v}
    + 9 \mu \mu_5 \frac{e \mk\hbar}{\kk\mk \eta\mk \pi^4\mk T_0^2} \left(\vec{E} +  \vec{v} \times \vec{B} \right)
    + \frac{3\mk e\mk\hbar}{2\pi^2 T_0^2} \mu \bb
    + \frac{\hbar}{2 T_0} \left(1+
    \frac{3\mu^2 + 3\mu_5^2}{\pi^2}\right) \left(\vec{\omega} + \vec{v} \times  \pd_\ttt \vec{v}\right)
    \right]
    \\
    = -\gf \mu_5\left(1 + \frac{\mu_5^2 + 3 \mu^2}{\pi^2}\right)
    + \frac{3 e^2 \hbar}{2\pi^2 T_0^3} \mk \vec{E} \cdot \vec{B} \,,
    \label{eq:mu51}
\end{multline}
\end{widetext}
where we have assumed the (rescaled) temperature $T_0$ to be independent of time and space. There would be additional effects from temperature inhomogeneities and fluctuations, but these are beyond the scope of this paper. See \Cite{ref:armstrong26} where these effects are shown to induce additional heating in the plasma at second order.

The second term in the first line of \cref{eq:mu1} is obtained by noting that the temporal component of the electric current [see \cref{eq:jterm1,eq:jterm2}] can be simplified by considering \(\jj \cdot \vv\) in \cref{eq:jj}, substituting it into the first line of \cref{eq:mu1} and neglecting \(\order{\vv^2}\) terms. In \cref{eq:mu1}, the first term in the second square bracket signifies the electric charge flow, the second term arises from the ohmic current, the third term represents the CME and the last term represents the CVE.
In \cref{eq:mu51}, the first term in each of the square brackets arises from the axial charge flow, and the second, third and fourth terms represent the CESE, CSE and AVE currents, respectively.

\Cref{eq:elecfield,eq:ind1,eq:k5,eq:mom,eq:mass,eq:mu1,eq:mu51} constitute the full set of chiral MHD 
equations, in the limit of subrelativistic bulk velocity.
The high number of terms in \cref{eq:mu1,eq:mu51} reduces significantly in the limit of small chemical potentials, to which we turn next.


\subsection{Small chemical potential limit}
\label{eq:mainchiraleqs}

These equations can be further simplified greatly by assuming that \(\mu\) and \(\mu_5\) are sufficiently small such that
\begin{equation}
\begin{gathered}
    \mu^2, \mu_5^2 \ll 1 \,,
    \quad
    \mu\mu_5 \pd_\ttt \mu_5 \ll \pd_\ttt \mu \,,
    \\
    \mu\mu_5 \pd_\ttt \mu \ll \pd_\ttt \mu_5 \,,
    \\
    \mu\mu_5 \norm{\del \mu_5} \ll \norm{\del \mu} \,,
    \quad
    \mu\mu_5 \norm{\del \mu} \ll \norm{\del \mu_5} \,,
    \\
    \hbar \qty(\vec \omega \cdot \vv) \pd_\ttt \qty(\mu^2 +\mu_5^2) \ll T_0 \pd_\ttt \mu_5 \,,
    \\
    \norm{\del \qty(\mu^2+3\mu_5^2) \times \vv} \ll \norm{\del \times \vv} \,.
\end{gathered}
\label{eq:smallchem}
\end{equation}
These conditions typically hold in the early Universe \cite{ref:domcke23,ref:brandenburg24,ref:gurgenidze26} and in heavy-ion collisions \cite{ref:kharzeev24}. With this, \cref{eq:mu1,eq:mu51} reduce to
\begin{multline}
    \pd_\ttt \left[
    \mu
    + \frac{3e\hbar}{2 \pi^2 T_0^2\kk} \vv \cdot \qty(\del \times \bb)
    \right]
    \\
    + \del \cdot \left[
    \mu \vec{v}
    + \frac{3e\hbar}{2 \pi^2 T_0^2\eta \kk} \left(\vec{E} +  \vec{v} \times \vec{B} \right)
    \frac{}{}\right.\\ \left.
    + \frac{3e\hbar}{2 \pi^2 T_0^2} \mu_5 \vec{B}
    + \frac{3\hbar}{\pi^2 T_0} \mu \mu_5 \left(\vec{\omega} + \vec{v} \times  \pd_\ttt \vec{v}\right)
    \right] = 0 \,,
    \label{eq:mu2}
\end{multline}
\begin{multline}
    \pd_\ttt \left[
    \mu_5 
    + 9 \mu \mu_5 \frac{e \hbar}{\kk \eta \pi^4 T_0^2} \vec{v} \cdot \vec{E}
    + \frac{3 e \hbar}{2 \pi^2 T_0^2} \mu \vec{v} \cdot \vec{B}
    \right] \\
    + \frac{\hbar}{T_0} \vec{\omega} \cdot \pd_\ttt \vec{v}
    + \del \cdot \left[
     \mu_5 \vec{v}
     + 9 \mu \mu_5 \frac{e \hbar}{\kk \eta \pi^4 T_0^2} \left(\vec{E} +  \vec{v} \times \vec{B} \right)
     \right.\\ \left.
    + \frac{3e\hbar}{2\pi^2 T_0^2} \mu \vec{B} \right]
    + \frac{3\hbar}{2\pi^2 T_0} \left(\vec{\omega} + \vec{v} \times  \pd_\ttt \vec{v}\right) \cdot \del \left(\mu^2 + \mu_5^2\right) 
    \\
    =  -\gf \mu_5 + \frac{3 e^2 \hbar}{2\pi^2 T_0^3} \mk \vec{E} \cdot \vec{B} \,,
    \label{eq:mu52}
\end{multline}
where we use \(\del \cdot \qty(\vv \times \pd_\ttt \vv) =  \vec \omega \cdot \pd_\ttt \vv - \vv \cdot \pd_\ttt \vec{\omega}\). Similarly, \cref{eq:ind1} simplifies to
\begin{multline}
    \frac{\pd \vec{B}}{\pd \ttt} = \del \times \left\{\vec{v} \times \vec{B}
    - \eta
    \left[ \del \times \vec{B}
    - \kk \mu_5 \vec{B}
    \frac{}{}\right. \right.
    \\ \left. \left.
    - \frac{2 \pi^2 T_0^2 }{3e \hbar} \kk \mu \vec{v}
    - \frac{2 T_0}{e} \kk \mu \mu_5 
    \left( \vec{\omega} + \vec{v} \times  \pd_\ttt \vec{v} \right) \right] \right\} \,.
    \label{eq:ind2}
\end{multline}
In order to simplify further, let us introduce two dimensionless variables
\begin{gather}
    \cm \equiv \frac{3 e \hbar \sqrt{\mu_0 \rho_0}}{2 \pi^2 T_0^2} \,,
    \qquad
    \cw = \frac{6\alpha}{\pi^3} \,,
    \label{eq:cmcv}
\end{gather}
where \(\rho_0\) is some energy density used to normalize the EM fields,
\begin{gather}
    \frac{\bb}{\sqrt{\mu_0 \rho_0}} \to \bb' \,,
    \quad \frac{\ee}{\sqrt{\mu_0 \rho_0}} \to \ee' \,,
    \label{eq:b_norm}
\end{gather}
and can be set to the background energy density as a convenient choice. The normalized EM fields are dimensionless in these units, with the normalized magnetic field equal to the Alfv\'en speed defined by the field strength and $\rho_0$. With this, \cref{eq:mu2,eq:mu52,eq:ind2} reduce to
\begin{multline}
    \pd_\ttt \left[
    \mu
    + \frac{\cm}{\kk} \vv \cdot \qty(\del \times \bb')
    \right]
    \\
    + \del \cdot \left[
    \mu \vec{v}
    + \cm \mu_5 \vec{B}'
    + \frac{\cm}{\eta \kk} \left(\vec{E}' +  \vec{v} \times \vec{B}' \right)
    \right.\\ \left.
    + \frac{\cw}{\kk} \mu \mu_5 \left(\vec{\omega} + \vec{v} \times  \pd_\ttt \vec{v}\right)
    \right] = 0 \,,
    \label{eq:mu3}
\end{multline}
\begin{multline}
    \pd_\ttt \left[
    \mu_5 
    + \mu \mu_5 \frac{6\cm}{\pi^2 \kk\mk \eta} \vec{v} \cdot \vec{E}'
    + \cm \mu \vec{v} \cdot \vec{B}'
    \right]
    + \frac{2\alpha}{\pi \kk} \vec{\omega} \cdot \pd_\ttt \vec{v}
    \\
    + \del \cdot \left[
     \mu_5 \vec{v}
     + \mu \mu_5 \frac{6 \cm}{\kk \mk\eta\mk \pi^2} \left(\vec{E}' +  \vec{v} \times \vec{B}' \right)
    + \cm \mu \vec{B}' \right]
    \\
    + \frac{\cw}{2 \kk} \left(\vec{\omega} + \vec{v} \times  \pd_\ttt \vec{v}\right) \cdot \del \left(\mu^2 + \mu_5^2\right)
    \\
    = -\gf \mu_5 + \frac{2\kk}{\cw} \mk\cm^2 \, \vec{E}' \cdot \vec{B}' \,,
    \label{eq:mu53}
\end{multline}
\begin{multline}
    \frac{\pd \vec{B}'}{\pd \ttt} = \del \times \left[
    \vec{v} \times \vec{B}'
    - \eta
    \left( \del \times \vec{B}'
    - \kk \mu_5 \vec{B}'
    \right. \right. \\ \left. \left.
    - \frac{1}{\cm} \kk \mu \vec{v}
    - \frac{\cw}{\cm} \mu \mu_5 
    \left( \vec{\omega} + \vec{v} \times  \pd_\ttt \vec{v} \right) 
    \right)
    \right]
    \,,
    \label{eq:ind3}
\end{multline}
the electric field, velocity and density are given by
\begin{multline}
    \ee' =  -\vec{v} \times \vec{B}'
    + \eta
    \left[ \del \times \vec{B}'
    - \kk \mu_5 \vec{B}'
    - \frac{1}{\cm} \kk \mu \vec{v}
    \right. \\ \left.
    - \frac{\cw}{\cm} \mu \mu_5 
    \left( \vec{\omega} + \vec{v} \times  \pd_\ttt \vec{v} \right) \right] \,,
    \label{eq:elec}
\end{multline}
\begin{gather}
    {\frac{D \vec{v}}{D \ttt}}
    = \frac{\rho_0}{\rho} (\vec{\del} \times \vec{B}') \times \vec{B}'
    - \frac{\vec{\del} p}{\rho} + \frac{\vec{\del} \cdot (2 \nu \rho \vec{\msf{S}})}{\rho} \,,
    \label{eq:mom2}
    \\
    \frac{D\rho}{D \ttt} = - \rho \mk \vec{\del}  \cdot \vec{v} \,.
    \label{eq:mass2}
\end{gather}
Simplifying the equations further requires assuming characteristic scales for $\cm$, \(\qty|\jj|\), and $|\vec{\omega}|$, which is addressed in \cref{sec:early}. In \cref{eq:mu3}, the second term in the first line represents the charge density associated with the three transverse currents in \cref{eq:j} and can be significant compared to \(\mu\) in general. The second term in the second line of \cref{eq:mu3} leads to chiral magnetic waves \cite{ref:kharzeev11,ref:ikeda23}, whereas the last term in that line causes damping of those waves. Since \cref{eq:elec} gives only the quasi-static electric field, it cannot consistently be used to eliminate $\ee'$ from this damping term. Thus, to account for this term properly would require solving for the electric field dynamically, along with a broader revision of the MHD equations used here, which is beyond the scope of this article. The term in the third line is associated with the CVE, and can be significant only if a large amount of vorticity is produced, since \(\cw \approx 10^{-3}\).

Similarly, in \cref{eq:mu53}, the CESE terms (the second term in each square bracket) can be estimated properly only by treating the electric field dynamically and their magnitude relative to the other terms depends on the hierarchy of small parameters and cannot be fixed a priori. The third term in the first square bracket can be ignored for sufficiently small $\cm$, but not in general. The third term in the second square bracket, however, leads to chiral magnetic waves. The last term in the first line also cannot generally be ignored, especially since the chiral plasma instability does produce significant vorticity \cite{ref:rogachevskii17,ref:schober18,ref:schober22,ref:brandenburg17}.  The last term on the left-hand side of \cref{eq:mu53} is the AVE contribution, which is again expected to play a role only if significant vorticity is produced. Notably, \cref{eq:mu53} provides an expression for the coefficient in the last term on the right-hand side, which can be used to estimate the strength of the magnetic field produced by the chiral dynamo through the conservation of the total chirality plus magnetic helicity for negligible chirality flipping rate \cite{ref:joyce97,ref:boyarsky12,ref:long16,ref:rogachevskii17,ref:boyarsky21b,ref:adler69,ref:bell69,ref:brandenburg05},
\begin{gather}
    \expval{\mu_5} + \frac{\kk}{\cw} \cm^2 \expval{\vec{A}' \cdot \bb'} = {\rm const.} \,,
    \label{eq:helchirm}
\end{gather}
where \(\bb' = \del \times \vec{A}'\) and the coefficient of \(\vec{A}' \cdot \bb'\) is equal to half the coefficient of \(\ee'\cdot \bb'\) in \cref{eq:mu53}. If we assume maximally helical fields with a length scale of \(2/(\kk \mu_5)\), we can roughly estimate the rms magnetic field produced from an initial \(\mu_5\)
\begin{gather}
    B\sub{rms} \approx \frac{\mu_5 \sqrt{\cw}}{\cm\sqrt{2}} \,,
    \label{eq:brms}
\end{gather}
which is independent of \(\kk\), as expected on dimensional grounds. Note that the spatial average here is over a length scale larger than the characteristic scale of the chiral plasma instability rather than over the whole Universe, which means that even fluctuations of \(\mu_5\) can locally produce magnetic fields of the appropriate helicity \cite{ref:schober24prl,ref:schober24prd}.
 
From this, we can also estimate the magnetic Reynolds number for the linear phase if we assume equipartition,
\begin{gather}
    v\sub{rms} \approx B\sub{rms} \,,
    \label{eq:vrms}
    \\
    {\rm Re_M} \approx \frac{2 B_{\rm rms}}{\mu_5 \kk \eta}
    = \frac{\sqrt{2\cw}}{\cm \eta \kk} \,.
    \label{eq:rm}
\end{gather}
During the saturation phase of the dynamo, when the field shifts to larger length scales \cite{ref:brandenburg17,ref:schober18} by a factor of $f \approx \lambda^{\rm saturation}/\lambda^{\rm linear}$ while still satisfying \cref{eq:helchirm}, the magnetic Reynolds number \({\rm Re_M} \to \sqrt{f}{\rm Re_M}\).

In \cref{eq:ind3}, the second term in the square brackets leads to the chiral dynamo, whereas the third term, associated with the charge flow, can lead to additional amplification and evolution of the field. We study this term in more detail in the follow-up paper \cite{my:cflow_theory}, where we examine the linear instability this term leads to.
The last term in the square brackets represents an additional dynamo contribution from the CVE, which can also act as an anomalous battery, generating magnetic fields from initial vorticity \cite{ref:wang24,ref:das25,tex:wang25,ref:wang26,ref:yamamoto16}.
The charge-flow term discussed above can play a similar role.

\section{Application to the early Universe}
\label{sec:early}

\subsection{Estimates}

Until now, we have kept the discussion largely general, subject to the constraints of small $\vv$, $\mu$, and $\mu_5$. Let us estimate some of the coefficients discussed above, which will also allow us to simplify the equations further. 
Furthermore, since we will 
discuss temperature-dependent physics, we recall that \(T\) denotes the physical temperature of the plasma and $T_0$ its rescaled value. In general, we will denote present-day values with the subscript
$0$.

Let us first consider the case without the chirality-flipping term, which we discuss below in more detail. During the assumed radiation-dominated era, the background energy density \(\bar \rho\) and the Hubble parameter \(H\), both at temperature $T$, can be written as
\begin{gather}
    \bar \rho = \frac{\pi^2}{30 \hbar^3} g_\ast T^4 \,,
    \qquad H^2 = \frac{8 \mk\pi\mk G}{3} \bar \rho \,,
\end{gather}
from which we can calculate the rescaled energy density \(\rho_0 \equiv \bar \rho a^4\). Here, $g_\ast$ is the number of relativistic degrees of freedom. Note that $\rho_0$ is what enters the normalization of the previous section \eqref{eq:b_norm} and has the same rescaling as the energy density in the momentum equation \eqref{eq:rhoscaling}. Using this, we can evaluate our chiral MHD parameters,
\begin{gather}
    \sqrt{\mu_0 \rho_0} = 1.62 \sqrt{g_*} \SI{}{\mu G}
    \,, \quad\! \bar \bb = \SI{1.62}{\mu G} \qty(\frac{T}{T_0})^2 \sqrt{g_*} \bb' \,,
    \label{eq:maximal_field}
    \\
    \frac{\kk}{a H}
    = \frac{2 \sqrt{3} \mk\mk a \mk \alpha T_0}{\pi \hbar \sqrt{8\mk \pi\mk G \rho_0}}
    \approx \num{1.45e29} \frac{a}{\sqrt{g_*}} \,,
    \label{eq:k5perh}
    \\[10pt]
    \cm
    = \frac{3 \mk e\mk \hbar \sqrt{\mu_0 \rho_0}}{2 \pi^2 T_0^2}
    = \sqrt{\frac{3 \alpha g_*}{10 \pi}}
    \approx \num{2.64e-2} \sqrt{g_*} \,,
    \label{eq:cm}
    \\[10pt]
    \cw = \frac{6\alpha}{\pi^3} \approx \num{1.4e-3} \,,
    \label{eq:cw}
\end{gather}
where we used \cref{eq:scaling1,eq:b_norm} for \cref{eq:maximal_field}, and \cref{eq:k5} for \cref{eq:k5perh} along with the fact that the corresponding unscaled value of $\kk$ would be $\kk/a$.\footnote{We have used the present-day value of $\alpha$ although it could be slightly higher, depending on $T$. This yields additional \(\order{1}\) effects.} Although the ratio in \cref{eq:k5perh} declines toward earlier times, it nevertheless remains very large even at the EW scale ($a \simeq \num{e-15}$), from which it can be concluded that the scales of chiral MHD lie far inside the Hubble radius at all times and the effects of the Hubble expansion can be neglected. Another scale for comparison would be the resistive scale. If we use the estimates from \Cite{ref:caprini09}, \(\eta = \hbar \alpha/(4T_0)\), we obtain
\begin{gather}
    \eta \kk = \frac{\alpha^2}{2\pi} \,,
    \label{eq:etak5}
\end{gather}
which can be substituted into \cref{eq:rm} to obtain
\begin{gather}
   {\rm Re_M}
    \approx \frac{\sqrt{2\cw}}{\cm\eta\kk}
    \approx \frac{\num{e5}}{\sqrt{g_*}} \,,
\label{eq:rm_early}
\end{gather}
which is independent of \(\mu_5\). Recall that any 
magnetic field amplification
beyond the linear phase could also increase this estimate by a factor of $\sqrt{f}$ where $f$ is the ratio of the length scale after saturation to the linear instability scale. Substituting \cref{eq:cm,eq:cw} into \cref{eq:brms} also yields an estimate of the strength of the magnetic field,
\begin{gather}
    B\sub{rms} \approx \frac{\mu_5}{\sqrt{g_*}} \,,
    \label{eq:brms2}
\end{gather}
ensuring that the assumptions of small \(B\sub{rms}\) and small \(\mu_5\) are indeed compatible. Although a larger \(\mu_5\) yields a stronger \(B\sub{rms}\), it does not impact the magnetic Reynolds number, which in turn determines the separation between the instability and resistive scales. This is one motivation for keeping the definitions of $\mu_5$ and $\kk$ separate.

If we assume equipartition, \ie \(v\sub{rms} \approx B\sub{rms}\) and the kinetic length scale is \(2/\qty(\kk \mu_5)\), we can also get an estimate for the kinetic Reynolds number \({\rm Re} = {\rm Re_M}/{\rm Pr_M}\) where 
\({\rm Pr_M}\) 
is the magnetic Prandtl number, which was estimated in \Cite{ref:caprini09} to be
\begin{gather}
    {\rm Pr_M} \approx \num{e12} \qty(\frac{\SI{1}{GeV}}{T})^4
    \enspace \to \enspace
    {\rm Re} \approx \num{e-7} \frac{1}{\sqrt{g_*}} \qty(\frac{T}{\SI{1}{GeV}})^4 \,,
    \label{eq:Re}
\end{gather}
from which we obtain that \({\rm Re} \approx 1\) around the temperature \(T \approx \SI{100}{GeV}\), where we used $\sqrt{g_*} \approx 10$ near the EW scale. As we briefly discussed after \cref{eq:mom}, this dependence on the temperature scale comes from the temperature-dependence of the rescaled kinematic viscosity $\nu$. If the physical kinematic viscosity scaled as \(\nu^{\rm phys} \propto a \propto T^{-1}\) like $\eta^{\rm phys}$, there would be no temperature dependence of the Prandtl number or the rescaled Reynolds numbers. Note that since \(\eta^{\rm phys}\) and \(\nu^{\rm phys}\) are rescaled identically, the physical Prandtl number and the rescaled Prandtl number are the same.

\addvspace{3mm}
\noindent\textbf{\textit{Chirality-flipping term}}.\,---\, 
Now, let us consider the rescaled chirality-flipping rate \eqref{eq:gammaf}. As described in \cite{ref:boyarsky21,ref:boyarsky21b}, at temperatures below the EW scale, \(T 
\lesssim
\SI{100}{GeV}\), the physical flipping rate \(\gf^{\rm phys} \propto T^{-1}\), yielding
\begin{gather}
    \gf \approx T_0 \alpha \frac{m_e^2}{T^2 \hbar} = \frac{\pi}{2} \kk \qty(\frac{m_e}{T})^2 \,,
    \label{eq:gammaf_low}
\end{gather}
which is clearly dependent on the temperature scale. In order for the chiral instability to operate, this rate has to be smaller than the chiral dynamo growth rate, which peaks at \(\gamma _5\approx \eta \mu_5^2 \kk^2/4\) \cite{ref:joyce97,ref:boyarsky12,ref:rogachevskii17}, giving us an order-of-magnitude estimate of the minimum temperature scale for the chiral dynamo to be efficient, \(\gamma_5 > \gf\),
\begin{align}
    T> T\sub{min} =
     m_e \frac{2\pi}{\alpha \mu_5}
    \approx \frac{\SI{0.4}{GeV}}{\mu_5} \,,
    \label{eq:tmin}
\end{align}
where we used \cref{eq:etak5}. Since \(\mu_5\) is assumed to be small, for even a modestly small \(\mu_5 \approx 0.1\),  
\(T\sub{min} = \SI{4}{GeV} \). 
This is, of course, a very simple estimate based on the linear growth rate, neglecting nonlinear effects. At this temperature, from \cref{eq:Re}, we estimate the kinetic flow to be strongly laminar with ${\rm Re} \approx \num{e-5}/\sqrt{g_*}$. Even if the assumed length scales for the magnetic field turn out to be $\approx 10$ orders of magnitude larger, this would increase the Reynolds number by only 5 orders of magnitude (via the $\sqrt f$ scaling above), still not turbulent at $T \approx \SI{4}{GeV}$. This assumes that the initial condition contains nonzero $\mu_5$. See \Cites{ref:gurgenidze26,ref:skoutnev26} for a discussion on the source terms.

Above the EW scale, the physical flipping rate \(\gf^{\rm phys} \propto T\), from which we obtain the rescaled flipping rate
\begin{gather}
    \gf = \qty(\frac{T_R}{M_*}) \frac{T_0}{\hbar} \,,
    \qquad M_* = m_{Pl} \sqrt{\frac{45}{4 \pi^3 g_*}} \,,
\end{gather}
where \(m_{Pl} = \SI{1.22e19}{GeV} \) is the Planck mass, \(g_* = 106.75\) above the EW scale, and $T_R \approx \SI{85}{TeV}$ is the temperature at which \(\gf^{\rm phys} = H = T^2/(M_*\hbar)\). Putting all of this together for temperature scales above EW yields \(\gf \approx \num{2.57e-11} \kk\), which is independent of temperature and is equal to the low-temperature estimate \eqref{eq:gammaf_low} roughly at the EW transition temperature. Since this flipping rate is independent of temperature, the equation for \(\mu_5\) once again becomes independent of the scale factor and the Hubble expansion.

\subsection{Simplified equations}

Let us now use the estimates from the previous subsection to simplify the equations further for the application to the early Universe. 
We use \cref{eq:cm,eq:cw,eq:etak5,eq:brms2,eq:vrms} to simplify \cref{eq:mu3,eq:mu53,eq:ind3,eq:elec,eq:mass2,eq:mom2},
\begin{widetext}
\begin{gather}
    \pd_\ttt \left[
    \mu
    + \highlight{\frac{\cm}{\kk} \vv \cdot \qty(\del \times \bb')}
    \right]
    + \del \cdot \left[
    \mu \vec{v}
    + \cm \mu_5 \vec{B}'
    + \frac{\cm}{\eta \kk} \left(\vec{E}' +  \vec{v} \times \vec{B}' \right)
    + \frac{\cw}{\kk} \mu \mu_5 \vec{\omega}'
    \right] = 0 \,,
    \label{eq:mu4}
    \\
    \pd_\ttt \mu_5 
    + \highlight{\frac{\pi^2 \cw}{3 \kk} \vec{\omega} \cdot \pd_\ttt \vec{v}}
    + \del \cdot \left[
     \mu_5 \vec{v}
     + \mu \mu_5 C_{\rm CESE} \left(\vec{E}' +  \vec{v} \times \vec{B}' \right)
    + \cm \mu \vec{B}' \right]
    + \frac{\cw}{2 \kk} \vec{\omega}' \cdot \del \left(\mu^2 + \mu_5^2\right)
    \nonumber\\
    \hspace{10cm}
    = -\frac{\pi}{2} \kk \qty(\frac{m_e}{T\sub{f}})^2 \mu_5 + 0.987 g_* \mk \kk \mk \vec{E}' \cdot \vec{B}' \,,
    \label{eq:mu54}
\end{gather}
\begin{gather}
    \frac{\pd \vec{B}'}{\pd \ttt} = \del \times \left[\vec{v} \times \vec{B}'
    - \eta
    \left( \del \times \vec{B}'
    - \kk \mu_5 \vec{B}'
    - \highlight{C_{\rm flow} \kk \mu \vec{v}}
    - \frac{\cw}{\cm} \mu \mu_5 \vec{\omega}' 
    \right) \right] \,,
    \label{eq:ind4}
    \\
    \ee' =  -\vec{v} \times \vec{B}'
    + \eta
    \left( \del \times \vec{B}'
    - \kk \mu_5 \vec{B}'
    - \highlight{C_{\rm flow} \kk \mu \vec{v}}
    - \frac{\cw}{\cm} \mu \mu_5 
    \vec{\omega}' \right) \,,
    \label{eq:elec4}
    \\
    {\frac{D \vec{v}}{D \ttt}}
    = \frac{\rho_0}{\rho} (\vec{\del} \times \vec{B}') \times \vec{B}'
    - \frac{\vec{\del} p}{\rho} + \frac{\vec{\del} \cdot (2 \nu \rho \vec{\msf{S}})}{\rho} \,,
    \qquad\qquad
    \frac{D\rho}{D \ttt} = - \rho \mk \vec{\del}  \cdot \vec{v} \,,
    \label{eq:mass4}
    \qquad \qquad \vec{\omega}' = \vec{\omega} +  \vec{v} \times  \pd_\ttt \vec{v} \,,
\end{gather}
\end{widetext}

~\\ \\
the various constants used are
\begin{gather}
    \cm = \sqrt{\frac{3 \alpha g_*}{10 \pi}}
    \approx \num{2.64e-2} \sqrt{g_*} \,,
    \\
    C_{\rm flow} = \frac{1}{\cm} \approx \frac{37.9}{\sqrt{g_*}} \,,
    \\
    \cw = \frac{6\alpha}{\pi^3} \approx \num{1.4e-3} \,,
    \\
    C_{\rm CESE} = \frac{6 \cm}{\kk \eta \mk \pi^2} \approx \num{1.89e3} \sqrt{g_*} \,,
    \\
    \eta \kk \approx \num{e-5} \,,
    \label{eq:eta_k5}
\end{gather}
the temperature \(T\sub{f} = \min \{T, \, T_{\rm EW}\}\) in \cref{eq:mu54} is the energy scale of the background plasma or the EW scale, whichever is less. The terms that differ from the standard chiral MHD equations \cite{ref:rogachevskii17,ref:schober24prd} are marked in \highlight{red}. 

The new terms in \cref{eq:mu54,eq:mu4} arise from the temporal components of the transverse currents, \ie the associated charge densities.
They can become comparable to the respective chemical potentials when the vorticity is large, as may occur in a chiral dynamo, or when the electric current is large. Therefore, before dropping these terms, it would be necessary to either have a diagnostic to check that these terms remain small in a numerical study, or to state their smallness as an explicit assumption in analytical work. In particular, all the vorticity-dependent terms can be dropped if \(\cw \norm{\vec{\omega}'}/\kk \ll 1\). The new term in \cref{eq:mu4} can be understood as modifying the conservation of $\mu$. Integrating the equations leads to the conservation of the total $\mu$ plus the integral of the new term
\begin{gather}
    \expval{\mu} + \frac{\cm}{\kk} \expval{\vec{v} \cdot \qty(\del \times \bb')} = {\rm const.} \,,
    \label{eq:mucons}
\end{gather}
where $\expval{\ldots}$ denotes the spatial average. The second term here is similar in structure to cross helicity except the magnetic field is replaced by the electric current. Thus, in the presence of this ``cross-current-helicity," the conservation of $\mu$ can get altered significantly. Also note that this conservation with the new term is then similar in structure to the conservation of chirality plus magnetic helicity given by \cref{eq:helchirm}.

In a numerical analysis, \(\kk\) is a parameter that can be calibrated in several ways. The first method is by using the coefficient of \(\ee'\cdot\bb'\) in \cref{eq:mu54}. If the chemical potentials have units of wavenumber, \(\kk\) can be calculated by equating the typical value of the chemical potentials in the analysis to the assumed dimensionless potentials \(\mu_5^{\rm wavenumber} = \kk \mu_5^{\rm dimless}\), where \(\mu_5^{\rm dimless} = \mu_5^{\rm phys}/T\) is used in this work \eqref{eq:mu_norm}. So, for example, if the typical values of \(\mu_5\) used in such a simulation are around 20, and we assume that it corresponds to a chemical potential of \(\mu_5^{\rm phys}/T \approx \num{e-2}\), then \(\kk = 2000\). A third method, perhaps the least practical, is to solve \cref{eq:eta_k5} for $\kk$ using the value of $\eta$ employed in the simulation. Of course, from a theoretical standpoint, all of these methods should be consistent.
When consistent values of, for example, $\eta$ are not feasible in a numerical study, these different methods of calibrating $\kk$ quantify how far the numerical regime is from the theoretical values.

The remaining highlighted terms, in \cref{eq:ind4,eq:elec4}, are due to the charge flow. If we assume equipartition, this contribution can be as large as the chiral dynamo term when \(\mu \approx \cm \mu_5\). This term leads to a battery mechanism, generating magnetic fields from scratch if there is a non-vanishing chemical potential and velocity field since the new term itself does not depend on the magnetic field. Moreover, this term also leads to a linear magnetized plasma instability, which has not been discussed in the literature to the knowledge of the authors. This instability leads to amplification of the magnetic fields from vorticity, which itself is further amplified by the Lorentz force. Since the instability already occurs at the linear stage, we derive the growth rate analytically and compare it with numerical simulations in the follow-up paper \cite{my:cflow_theory}. These results, along with the modification detailed in \cref{eq:mucons}, underline the importance of including the new terms, which can alter the dynamics significantly.

\section{Conclusions}
\label{sec:concl}

To summarize, we present a systematic derivation of the equations of relativistic chiral MHD for a plasma of charged fermions in an expanding universe. Starting from the covariant Maxwell equations and the constitutive relations for the electric and axial currents, we show how the full system can be brought to the same form as in the Minkowski metric through a combination of a coordinate transformation and a rescaling of the dynamical variables, in direct analogy to the well-known procedure for ideal MHD.
Our presentation distinguishes explicitly between the coordinate transformation, which yields the physical quantities measured by a local observer, and the subsequent rescaling, which absorbs the scale factor.
The resulting equations are independent of the scale factor except through two well-identified channels: the chirality-flipping rate below the EW scale \eqref{eq:gammaf_low} and the rescaled kinematic viscosity, for both of which a dependence on the temperature scale survives the rescaling.

In deriving these equations, we retain all four contributions to both the electric and the axial current: the charge flow, the ohmic (respectively, CESE) current, the CME (respectively, CSE) current, and the CVE (respectively, AVE) current.
We use the limit of subrelativistic bulk velocity \eqref{eq:subrel} and small chemical potentials \eqref{eq:smallchem} to arrive at simplified equations \eqref{eq:mu3}--\eqref{eq:mass2}.
Our notation separates the dimensionless chemical potentials from the characteristic chiral wavenumber \(\kk\), which allows for differentiating the analysis of the length scales and the field strengths. For example, for the chiral dynamo, the produced magnetic field strength \(B\sub{rms}\) scales with \(\mu_5\) \eqref{eq:brms}, whereas the magnetic Reynolds number \({\rm Re_M}\) does not \eqref{eq:rm}. 

We consider the application to the radiation-dominated era in the early Universe, which allows us to simplify the equations and evaluate all the dimensionless coefficients \eqref{eq:mu4}--\eqref{eq:eta_k5}.
We find that, for the chiral dynamo, the resulting \({\rm Re_M} \approx 10^5/\sqrt{g_*}\) \eqref{eq:rm_early}, and \(B\sub{rms} \approx \mu_5\) \eqref{eq:brms2} in units of \(\sqrt{\mu_0 \rho_0/g_*} = \SI{1.62}{\mu G}\) (comoving), confirming that the assumptions of small \(\mu_5\) and small \(B\sub{rms}\) are mutually consistent.
Comparing the linear growth rate of the chiral plasma instability with the chirality-flipping rate yields a simple estimate for the minimum temperature for the operation of the chiral dynamo, \(T\sub{min} \approx \SI{0.4}{GeV}/\mu_5\) \eqref{eq:tmin}, at which the kinetic flow is expected to be laminar.
For numerical studies, we discuss methods for calibrating \(\kk\), and the conditions under which the vorticity-dependent and new charge-density terms are negligible.

The final equations include terms that, to our
knowledge, have not been considered in the chiral MHD literature. These are of two kinds. First, the temporal components of the transverse currents contribute charge densities that enter the time derivatives in the evolution equations for $\mu$ and $\mu_5$ [the highlighted terms in \cref{eq:mu4,eq:mu54}].
Second, the electric current associated with the flow of the charge density yields a term \(\propto C_{\rm flow}\mk \kk \mu \vv\) in the electric field and consequently in the induction equation [\cref{eq:elec4,eq:ind4}].
Notably, this charge-flow term is mandated by current conservation whenever \(\mu \neq 0\), requires no chirality imbalance, and thus persists even after \(\mu_5\) is fully depleted by the chiral dynamo or the chirality flipping.
For equipartition between magnetic and kinetic energy, it can exceed the CVE contribution and become comparable to the CME term, and it can also act as a battery, generating magnetic fields from an initial bulk flow.
In the follow-up paper \cite{my:cflow_theory}, we show that this term renders the system linearly unstable and study the resulting charge-flow instability in detail.

Several extensions remain for future work. The quasi-static closure for the electric field employed here cannot capture the damping of charge fluctuations through underdamped plasma oscillations, which is relevant both for the CESE terms and for the saturation of the C-flow instability \cite{my:cflow_theory}; a proper treatment requires retaining the displacement current, or ultimately a two-fluid description.
Furthermore, we assume the rescaled temperature to be homogeneous, and the effects of temperature fluctuations on the anomalous transport coefficients remain to be quantified. 
Finally, coupling the present framework to explicit sources of \(\mu\) and \(\mu_5\), such as those arising in baryogenesis scenarios and proto-neutron stars, would allow the new terms to be assessed self-consistently.

\begin{acknowledgments}
We thank B. Armstrong, A. Brandenburg, R. Durrer, A. Long, I. Rogachevskii, K. Seller, G. Sigl, and O. Sobol for helpful discussions.
\end{acknowledgments}

\appendix
\crefalias{section}{appendix}

\section{List of definitions and notation}
\label{sec:abbr}

For the reader's convenience, we list the abbreviations and symbols used in this work.

~\\
Abbreviations:
\begin{notation}[4em]
\item[CME] Chiral magnetic effect
\item[CVE] Chiral vortical effect
\item[CSE] Chiral separation effect
\item[CESE] Chiral electric separation effect
\item[AVE] Axial vortical effect
\item[EM] Electromagnetic
\item[MHD] Magnetohydrodynamics
\item[FLRW] Friedmann--Lemaître--Robertson--Walker
\item[EW] Electroweak
\end{notation}

~\\
Conventions:
\begin{notation}[4em]
\item[\(c \!=\! k\sub{B} \!=\! 1\)]
  units used throughout, with \(\hbar\) and \(e\) retained explicitly
\item[\(\epsilon_{\alpha\beta\gamma\delta}\)]
  totally antisymmetric Levi-Civita tensor with \(\epsilon_{0123} = \sqrt{-g}\)
\item[\(\bar X\)]
  physical quantity measured by a local (comoving) observer,
  obtained via transformation from FLRW to Minkowski
\item[\(\widetilde X\)]
  rescaled quantity absorbing the scale factor; tildes are dropped
  after \cref{sec:prelim}
\item[\(\vec \omega'\)]
  $\vec\omega + \vv \times \pd_\ttt \vv$, vorticity observed by the fluid element
\end{notation}

\newpage\noindent
Electric and magnetic fields:
\begin{notation}[4em]
\item[\(E^\alpha, B^\alpha\)]
  fields with respect to the comoving observer
\item[\(\bar\ee, \bar\bb\)]
  physical fields measured by a local observer
  \eqref{eq:ebbar1}
\item[\(\widetilde\ee, \widetilde\bb\)]
  rescaled fields, \(\widetilde\bb = a^2 \bar\bb\)
  \eqref{eq:scaling1}; tildes dropped after \cref{sec:prelim}
\item[\(e^\alpha, b^\alpha\)]
  fields observed by the fluid element \eqref{eq:emfluid}
\item[\(\ee', \bb'\)]
  rescaled fields normalized by \(\sqrt{\mu_0\rho_0}\) \cref{eq:b_norm}, where $\rho_0$ can be chosen to suit the problem
\end{notation}

~\\
Other symbols:
\begin{notation}[4em]
\item[\(a = a(\ttt)\)]
  scale factor of the FLRW metric \eqref{eq:metric}
\item[\(\ttt\)]
  conformal time
\item[\(n^\alpha\)]
  velocity of a comoving observer,
  \(= (1/a, 0, 0, 0)\)
\item[\(u^\alpha\)]
  bulk velocity of the fluid,
  \(u^\alpha = \gamma (1, \vv)/a\)
\item[\(\vv\)]
  bulk three-velocity of the fluid; \(v = \abs{\vv}\)
\item[\(T\)]
  physical (unscaled) plasma temperature
\item[\(T_0\)]
  rescaled plasma temperature; for early-Universe application: present-day CMB value
\item[\(\mu\) ($\mu_5$)]
  dimensionless (chiral) chemical potential,
  \\
  $= \mu^{\rm phys} (\mu_5^{\rm phys})/T$
\item[\(\rho_{\rm el}, \rho_5\)]
  electric and axial charge densities, with units of number density multiplied by $e$
\item[\(\alpha\)]
  fine-structure constant, \(= e^2 \mu_0/(4\pi\hbar)\)
\item[\(g_*\)]
    number of relativistic degrees of freedom
\item[\(\kk\)]
  characteristic wavenumber of the chiral dynamics, $= 2\alpha T_0/(\pi\hbar)$; the chiral
  plasma instability grows maximally at \(k_5 = \kk\mu_5/2\)
\item[\(\cm\)]
  CME/CSE coefficient, \(= 3 e \hbar \sqrt{\mu_0\rho_0}/(2\pi^2 T_0^2)\)
\item[\(\cw\)]
  CVE/AVE coefficient, \(= 6\alpha/\pi^3 \approx \num{1.4e-3}\)
\item[\(C_{\rm flow}\)]
  charge-flow coefficient, $= 1/\cm$
\item[\(C_{\rm CESE}\)]
  CESE coefficient, \(= 6\cm/(\kk\eta\pi^2)\)
\item[\(\gf\)]
  (rescaled) chirality-flipping rate
\item[\(T\sub{f}\)]
  temperature \(T\sub{f} = \min\{T, T_{\rm EW}\}\) in \cref{eq:mu54}
\end{notation}

%

\end{document}